\documentclass[12pt]{article}

\usepackage{scicite}
\usepackage{times}
\usepackage{graphicx}
\usepackage{amsmath}
\usepackage{xcolor}
\usepackage{times}
\usepackage{url}

\newenvironment{sciabstract}{%
\begin{quote} \bf}
{\end{quote}}

\title{Scalable photonic diffractive generators through sampling noises from scattering medium}

\author
{Ziyu Zhan,$^{1}$ Hao Wang,$^{1}$ Qiang Liu,$^{1,2,3\ast}$ Xing Fu$^{1,2,3\ast}$ \\
\\
\normalsize{$^{1}$Department of Precision Instrument, Tsinghua University,}\\
\normalsize{Beijing 100084, China}\\
\normalsize{$^{2}$State Key Laboratory of Precision Space-time Information Sensing Technology, }\\
\normalsize{Beijing 100084, China}\\
\normalsize{$^{3}$Key Laboratory of Photonic Control Technology (Tsinghua University),}\\
\normalsize{Ministry of Education, Beijing 100084, China}\\
\normalsize{$^{\ast}$To whom correspondence should be addressed:}\\
\normalsize{fuxing@tsinghua.edu.cn; qiangliu@tsinghua.edu.cn}\\
}

\date{}

\begin{document} 


\baselineskip24pt


\maketitle 
\begin{sciabstract}
Photonic computing, with potentials of high parallelism, low latency and high energy efficiency, have gained progressive interest at the forefront of neural network (NN) accelerators. However, most existing photonic computing accelerators concentrate on discriminative NNs. Large-scale generative photonic computing machines remain largely unexplored, partly due to poor data accessibility, accuracy and hardware feasibility. Here, we harness random light scattering in disordered media as a native noise source and leverage large-scale diffractive optical computing to generate images from above noise, thereby achieving hardware consistency by solely pursuing the spatial parallelism of light. To realize experimental data accessibility, we design two encoding strategies between images and optical noise latent space that effectively solves the training problem. Furthermore, we utilize advanced photonic NN architectures including cascaded and parallel configurations of diffraction layers to enhance the image generation performance. Our results show that the photonic generator is capable of producing clear and meaningful synthesized images across several standard public datasets. As a photonic generative machine, this work makes an important contribution to photonic computing and paves the way for more sophisticated applications such as real world data augmentation and multi modal generation.
\end{sciabstract}

\section*{Introduction}
Neural networks (NN) have shaped the landscape of machine intelligence by revolutionizing the way computers perform functions and make decisions\cite{lecun2015deep}. Such advances are certainly underpinned by the performance improvements of silicon-based digital processors over the past few decades\cite{shekhar2024roadmapping}. Nevertheless, a noticeable disparity has emerged between the capability to process larger volumes of data in a more efficient and faster manner and the limited scalability of digital circuits in the current era\cite{lu2014toward,mehonic2022brain}. This unbalance has consequently echoed the interest of engineering analog optical computers under the umbrella of non-von Neumann architecture\cite{furber2016large,schuman2017survey,markovic2020physics,chen2023all}. Through physically integrating data processing and storage, breaking down the barrier between hardware and software, and harnessing the high parallelism, energy efficiency and low latency enabled by optical waves, ubiquitous neuromorphic photonic frameworks are invented as special-purpose accelerators\cite{shastri2021photonics,mcmahon2023physics,wetzstein2020inference}. For instance, the large bandwidth of photonic along with maturing fabrication techniques is exploited to demonstrate an in-memory photonic tensor core capable of convolutionally processing images\cite{xu202111, feldmann2021parallel,chang2018hybrid,ashtiani2022chip}, large-scale diffraction photonic neural networks to achieve multi-layer perceptrons\cite{bueno2018reinforcement,lin2018all,zhou2021large,chen2023photonic}, and on-chip large coherent photonic networks to efficiently realize linear matrix-vector multiplication\cite{shen2017deep,hughes2018training}. Beyond these linear operations, nonlinear activation functions are gaining efforts as well, hoping to implement deeper NN architectures in photonic\cite{wang2023image}. Critically, though these findings suggest that discriminative optical NNs with classification capabilities are within reach, photonic generative NNs remains largely unexplored. 

Generative models, such as ChatGPT, are designed to learn and replicate the abstract patterns or structures hidden in the data and are expected to produce novel and realistic outputs after optimization\cite{goodfellow2020generative,creswell2018generative}. These models have supported diverse applications including image synthesis, text generation, data augmentation among many others\cite{zhang2018stackgan++, fedus2018maskgan,reed2016generative,antoniou2017data}, and have become a milestone in the pursuit of artificial general intelligence. Unlike training discriminative networks, which comparably  concentrate on learning boundary between classes within a dataset, many generative models center on learning the distribution of the classes and usually require an additional data sampler. Sampling is crucial here because it determines the model's ability to generate new data points aligning with the original data statistics\cite{white2016sampling}. The need for a data sampler, along with complicated training, results in the challenging photonic implementation of such generators. Notably, a pioneering design of a photonic generative network incorporates a physical random number generator as the data sampler and phase-change metasurfaces as computation weights. Though being inspiring, the sampler is strictly required to convert random optical noises as digital voltages to realize randomness, while optoelectronic NN is heavily established on a $2\times2$ tensor core. The switch among domains of the sampler and the network greatly hinders an efficient and hardware-consistent implementation. The limited scale of computation weights constrains the generator's expressivity, allowing it to generate only a single digit\cite{wu2022harnessing}.  

In this work, we exploit the multiple light scattering as a natural noise source and present a scalable photonic diffractive generator (PDG) with $100\times 100$ programmable weights capable of generating multiple digits' images to achieve over $10^{8}$ optical computing operations. By pursuing the spatial parallelism of light, we break the boundary between the optical noise sampler and the PDG thus demonstrate a more powerful generative network than achieved previously. Our PDG elaborately manipulates the spatial modes of light through diffractive phase layers coupled with free-space propagation. Regarding the noise source, we experimentally collect the scattering responses of a disordered medium at many spatial illuminating positions. Prior to this, we design two encoding strategies (random encoding and physics-aware encoding) that compress images to the illuminating position coordinates, which serves as the physical latent space of the noise sampler. The obtained speckle patterns and images are formed as pairs to optimize subsequent photonic generative NN, assisted by another digital discriminative network using generative adversarial network (GAN) framework. After training, through experiments and simulations, we validate PDG can produce clear and diverse images with different complexities thanks to the proposed encoding methods and scalable NN architectures. Last, we demonstrate that the generator can achieve image interpolation by simply varying illumination position of the incident light without any further training or processing.

\section*{Results}
\subsection*{Overall framework}
A schematic of the our framework is illustrated in Fig. 1. In the proposed GAN, there are 3 essential blocks including a photonic noise sampler, photonic diffractive generator $\boldsymbol{G}$ and a digital discriminator $\boldsymbol{D}$ (Fig. 1A). Generally, the generator takes a noise input to output a candidate `fake' image $\boldsymbol{o}$, and the discriminator endeavors to distinguish between `fake' and `real' data. During training, these two models engage in adversarial interactions and are believed to reach a Nash equilibrium state upon convergence. Subsequently, the generator is ready to produce new vivid data for inference. To achieve random noises in hardware, we exploit the light scattering in a disordered medium (ground glass) with inherent randomness (Fig. 1B). Such a complex process involves numerous interference events that give rise to a speckle pattern and can be described by a linear transmission matrix with Gaussian i.i.d elements. 

Though this scattering effect is seemingly detrimental for applications like imaging, the intrinsic statistical properties can be tamed as an appropriate natural random noise generator, especially considering that GANs are probabilistic models that learn to generate samples by capturing the underlying probability distribution of the training data. But to use this noise as meaningful input to the generator, pre-determined  link between certain speckle patterns (noise input) and images (output) is required before training. 

We therefore propose two encoding methods, i.e. random position encoding and physics-aware position encoding (Fig. 1C). For both of them, the lateral illuminating positions $\boldsymbol{v}=(x, y)$ of the scattering medium are essentially interpreted to the coordinates of latent parameter space of physical noise sampler. Given $\boldsymbol{v}$, a deterministic noise vector $\boldsymbol{z}=\boldsymbol{S}(\boldsymbol{v})$ reshaped from a two-dimensional speckle pattern can be acquired. The difference between them is that the former associates an image with a random position $\boldsymbol{v}$ while physics-aware encoding exploits a trained encoder to designate an image to the position coordinates (see Supplementary Text 2). To pursue the rich spatial modes of speckle patterns, we implement the generative part adopting diffractive deep neural networks for the end-to-end hardware consistency and for the sake of a potential all-optical implementation (Fig. 1D). The combination of multiple programmable optical layers used for light field modulation and fixed light propagation used for information mixing delivers rich computations and has empowered many applications. In experiments, we set up a reconfigurable optoelectronic block ( known as diffractive processing unit) with off-the-shelf components, e.g. spatial light modulators (SLM) and cameras (Fig. 1E). For each unit, the incident beam first undergoes computational amplitude and phase modulations and then propagates in free space, after which the extracted feature is recorded by a camera, which can be considered as a quadratic nonlinear activation function on the complex optical field. By introducing into more analog-digital and digital-analog conversions, such programmable setup allows us to attain comparable or even better results with diffractive deep NNs.  In a nutshell, the overall inference phase can be formalized as $\boldsymbol{o}=\boldsymbol{G}(\boldsymbol{S}(\boldsymbol{v}))$.

\subsection*{Photonic noise sampler}
To explain the random position encoding and physics-aware encoding in detail, we present comparison results in Fig. 2. Intuitively, a focused coherent laser illuminates a specific part of the scattering medium denoted by coordinates $\boldsymbol{v}=(x, y)$ and generates a speckle pattern as a result. To train the generator, an illuminating position, a resulting speckle pattern and a ground-truth image are bonded as pairs. To acquire adequate pairs, we scan across the medium by modulating incident light with various wavevectors (see Methods). Stated differently, the speckle patterns corresponding to these positions are fed as generator input and the paired images are considered as ground truths. For random encoding, we map images from a given dataset arbitrarily to these positions.  Instead of random designation, physics-aware encoding aims to transform similar noise inputs to similar image outputs in a logical way. To this end, we utilize another digital variational autoencoder (VAE) to firstly compress an image to a latent space — a 2-dimensional vector representing a position coordinate — and map the image to the resultant speckle (see Fig. 2A and details in Supplementary Text 2). Thanks to memory effect\cite{judkewitz2015translation}, when incident light is tilted or shifted a bit, the speckles tend to be correlated or slightly changed, which more or less corresponds to the smooth latent space of the trained encoder (thus `physics-aware'). An experimentally noise intensity pattern is shown in Fig. 2B. The fitted probability density function well approximates a Rayleigh distribution with experimental errors, as a result of Gaussian distribution of real and imaginary elements (see more theoretical discussions in Supplementary Text 1). The final mapping from the physics-aware encoder used in experiments is presented in Fig. 2C. As expected, the smooth trend of images are preservered by the encoder, manifested by the fact that neighboring images are changing gradually.

To compare them, we simulate two four-layer PDGs respectively. We use the image metric to evaluate the modal performance, namely Pearson correlation coefficient (PCC)\cite{cohen2009pearson, benesty2008importance} as image quality metric. The metric is a part of the overall training loss function, which evaluates the image quality of synthesized data to ground truths.  As shown in Fig. 2E, the generator trained with physics-aware encoder exhibits higher baseline, i.e. higher quality images than that of random encoding (see inset Fig. 2E) with the improvement of average PCC as $ \delta PCC = 0.151$.

\subsection*{Photonic diffractive generator}
To establish advanced generative network architectures in experiments, we implement cascaded and parallel diffractive layers built on diffractive processing unit respectively (Fig. 3). The cascaded architecture corresponds to the deep diffractive NN achieved previously, in which the output from one layer is fed into the next layer. As comparison, inspired by the broad learning approach in machine learning, we propose a purely parallel architecture of photonic NN (Fig. 3B). It extracts different features individually by each layer and then synthesizes all features together by a weighted summation to formalize the network output.  In this work, we exploit solely \textit{in-silico} co-training methods for both photonic generator and digital discriminator and perform the experimental inference phase for the photonic generator. The experimentally-implemented phase plates are shown in Fig. 3C. With the deliberated architecture and physic-aware encoding method, the PDG is demonstrated to experimentally generate all categories of hand-written images from MNIST (see Supplementary Figure 2 for setup calibration and Supplementary Figure 4 for sequence diagram of timing control). Additionally, to enhance the image generation quality, we explore the use of a linear digital layer instead of optical readout from the camera. This greatly increases the image quality thanks to high precision operation of digital computers (Fig. 3D). We summarize the experimental results in a scatter plot in Fig. 3E. 

Furthermore, to investigate the image generation capability of the proposed PDG, we challenge it with more complex datasets including Fashion-MNIST and EMNIST in simulations. As illustrated in Fig. 4, a firm conclusion is that digital readout layer can improve the image quality while which architectures provide better results is more subtle and can depend on the target datasets.  We therefore posit that sophisticated hybrid architectures including parallel and cascaded submodules could be optimized for a best task-specific performance, as recently demonstrated in photonic discriminative networks.

Besides image generation, image interpolation is another fascinating generative task which has been used in data augmentation, artistic exploration, or semantic manipulation. It aims to create new data points that lies between two or more existing samples smoothly and continuously. For example, if a generative model has been trained on two different face profiles, selecting an appropriate trajectory between these corresponding inputs allows the generation of an intermediate turning-face process. For the photonic implementation, thanks to the memory effect of scattering media and the proposed sampling method, we can realize image interpolation conveniently by just scanning from one input position $\boldsymbol{v_1}=(x_1, y_1)$ to another position $\boldsymbol{v_2}=(x_2, y_2)$. The typical interpolation results on the MNIST dataset are shown in Fig. 5B. In this demonstration we employ a linear scanning of the scattering media from $\boldsymbol{v_1}$  to $\boldsymbol{v_2}$, while in practice, one could determine an optimal path connecting the input instances in the representation space.

\section*{Discussion}
We have demonstrated that the photonic generator with proposed sampling methods can generate images in a variety of settings. In this work, we provide a compelling case for implementing photonic generative network by effectively communicating the SLM and the camera via a computer which allows for flexible layer scaling in a hardware-efficient manner, while the fully-optical approaches can be potentially implemented via diffractive deep neural network with improved energy efficiency and system latency in future work. For both the cascaded and parallel architectures, we perform ablation study on the layer number of the network (see Supplementary Figure 3). As expected, more layers lead to performance improvement for both of them, which can attribute to the larger field of view (FOV) of the whole optical computing system\cite{rahman2022universal}. Though cascaded architectures reach a higher baseline than parallel counterparts from this simulation study, in realistic experiments, the system errors aggregates more severely for cascaded networks, which can result in the model collapse. Indeed, each architecture has its own advantages and disadvantages. Beyond this performance differences, cascaded networks is more physically interpretable as the speckle images are processed and transformed gradually layer-by-layer, while for parallel architectures, intermediate feature maps are still speckle-like and are lack of intuitive physical meaning. But the parallel PDG could be suitable for parallel timing control to significantly improve the data throughout rate. 

Inspired by the pipeline programming scheme widely used in field programmable gate arrays (FPGAs)\cite{zhang2015optimizing} and graphics processing units (GPUs)\cite{krizhevsky2012imagenet}, we can divide the total computing process into two modules: optoelectronic operation and host computer operation. Both of these modules are executed simultaneously and synchronized by system clock and mutual exclusion (Mutex) with the help of multi-thread programming scheme (see schematic details of parallel timing control in Supplementary Figure 7). For example, for a four-layer parallel PDG with digital readout layer, time consumption of two modules including computation and device communication tends to be at the same scale with 97.9 ms and 85.4 ms respectively, indicating that the parallel accelerating programming scheme will positively double the output data rate for PDG.

Though the noise input can be straightforwardly obtained by illuminating the scattering media, the scanning range is constrained by the system aperture experimentally. Moreover, the scanning step cannot be too small, especially smaller than the speckle size, since two adjacent optical speckles might be too similar due to the memory effect to be distinguishable by the following generator. Consequently, there exists an upper bound for the number of independent physical noise sources that can be sampled (a detailed mathematical analysis is provided in Supplementary Text 5). This upper bound can affect the sampling process for PDG, especially when dealing with complex datasets such as ImageNet. A possible solution to enhance the sample capacity is to replace the phase gratings used for scanning with specially designed grayscale patterns. Unlike the previous binary-like scanning method, this grayscale wavefront shaping strategy offers additional channels for noise sampling.

Based on the physics-informed encoding method, our PDG model is not just a photonic counterpart of a simple digital GAN, but a VAE-GAN combined model (see additional comparison between digital VAE-GAN between PDG in Supplementary Figure 6). Such combined architecture is an important fraction in recent generative model researches\cite{bao2017cvae,gao2020zero}, whose main idea is to replace the conventional element-wise error loss by such competing training strategy, meanwhile, the front-end VAE provides a relative smooth parameter space for back-end GAN. 

For the whole generator, we use a rough-tuned VAE-encoder to generate two-dimensional coordinates, and then randomly project them to the high-dimension complex signals by the optical scattering media. These noisy speckle patterns are finally processed by photonic NN. \textit{In other words, PDG mathematically acts stricto sensu a conventional complex GAN but harnesses a physics-informed encoding method based on prior knowledge from dataset.} However, we here just arbitrarily map images with certain spatial positions without considering any feedback knowledge of follow-up networks. This algorithm seems not to reach a global optimum for the whole encoding task. To attain a global optimum, we need to incorporate the VAE into the training process collectively. Due to the physical complexity and instability of scattering media, it is hard to execute accurate backpropagation model for a sufficiently long time. Fortunately, recent studies have proposed several adaptive solutions to tackle physical systems with noise, such as forward-forward algorithm \cite{hinton2022forward}, Bayes optimization\cite{snoek2012practical}, dual adaptive training\cite{zheng2023dual}, and physics aware training\cite{wright2022deep}. By utilizing above in-situ in-silico hybrid optimization algorithms, a joint training scheme could enhance the performance of PDG.

In summary, we have demonstrated a photonic noise sampler and a photonic diffractive generator to facilitate the implemention of generative networks in GANs. Through experiments and simulations, we substantiate that our system is suitable at producing meaningful images and seamlessly interpolating to novel data. Notably, we design two noise sampling methods to solve the generator training problem and to enhance the overall generation performance. The proposed PDG exhibits scalability and flexibility among various datasets, standing as an effective example with promising applications in image synthesis, anomaly detection and data augmentation in the realm of generative models.

\section*{Materials and Methods}
\subsection*{Experimental setup}
The detailed experimental setup to realize the PDG is illustrated in Supplementary Figure 1. We construct the noise sampler by employing a phase-only SLM (SLM-P1, $6 \ \mu m$ pixel pitch), an objective lens (1, NA=0.3), and a scattering medium (LBTEK, DW110-220). We form the DPU by utilizing an amplitude-only SLM (SLM-A, $8 \ \mu m$ pixel pitch UPOLabs, HDSLM80RA), a phase-only SLM (SLM-P2, $8\ \mu m$ pixel pitch, UPOLabs, HDSLM80R), and a sCMOS camera (PCO panda 4.2). We configure L3, L4 and L3, L6 as two 4f systems with two different apertures (AP1 and AP2). These two apertures are used for optical path selection. To direct light through path1, we load a phase grating with grating period ($\Delta_1$,0). The modulated light is filtered by AP2 with only first-diffraction entering while being blocked by AP1. Equally, to direct light through path2 for speckle generation, we will load a new phase grating with ($\Delta_x$,$\Delta_2+\Delta_y$), where $\Delta_x$ and $\Delta_y$ are scanning phase gratings and $\Delta_1 \approx \Delta_2 \approx \Delta_x \approx \Delta_y$. This allows us to choose the desired optical paths by simply adding corresponding phase masks on SLM-P1. It's worth noting that not using a galvanometer for scanning operation is because that a single galvanometer can hardly scan with such a small scanning step precisely. Though one could apply a telescope system at the output end of the galvanometer to enhance the angular resolution of scanning, it is challenging to incorporate a galvanometer system with the follow-up experiment devices. For cascaded PDG architecture, we first direct the light through path2 and enter DPU for front-end computing. Then we change the phase mask of SLM-P1 to let the light pass through path2 leading to a basic DPU functionality for the remaining layers. Specifically, the feature map of the penultimate layer can be fed into a shallow digital layer rather than the direct camera readout to improve the model performance. As for parallel PDG architecture, with the input optical path fixed on path1, individual featured maps are all stored in the host computer and computed for a synthesized feature map by a weighted summation of existing feature maps, which could be post-processed either electronically or optically.

Phase gratings with grating period ($\Delta x$,$\Delta y$) on a phase-only spatial light modulator (SLM) focus the modulated light by an objective lens with numerical aperture (NA=0.3) onto the scattering media. The position shifts for focus spots of different phase gratings are calculated as ($\Delta x$,$\Delta y$)=($\lambda f/\Delta x$, $\lambda f/\Delta y$), where $\lambda$ and $f$ are the wavelength and focus length of objective lens respectively. By successively loading different phase gratings, we can scan the light across scattering media to generate optical random signals. 

\subsection*{Data processing and figure evaluation metric}
For DPU experimental setup, we employ an amplitude-only SLM with pixel size of $8 \ \mu m $ to encode the input data with the range of $[0, 1] $ , where each input dimension is encoded by a macropixel with the size of $5\times 5$ . The overall encoded region occupies a region of $ 500 \times 500$ pixels at the central of the SLM plane and pixels in the rest of region are set to 0. We establish a 4f imaging system with two identical plano-convex lenses to map input encoding plane to phase modulation plane.  We flip both sides of the phase plates due to the reverse replica of Fourier transform in 4f imaging system and then load them to a phase-only SLM with the same pixel size of $8 \ \mu m$ for pixel-wised phase modulation. Notably, both SLMs process 10 bit-depth modulation ranging from 0 to 1023. In order to match the common HDMI communication protocol, we successively separate 10-bit number from higher to lower place into three parts with each bit-length as 3, 3, and 4, and send them to SLM controller via the red, green and blue channels in HDMI cable, respectively.
We set a region of interest ($615 \times 615$) of detection camera and then downsample the speckle images reducing the dimension to $100\times100$. For intermediate layer in cascaded architecture, the resultant feature maps need to be resized to a scale of  $500\times500$ with nearest interpolation techniques. As for optical readout, we utilize a Gaussian filter with size of $12\times 12$ in front of final output to alleviate the noise in photoelectric conversion. 
Here we introduce a figure evaluation metrics called Pearson Correlation Coefficient (PCC) as training loss function which is defined as:
\begin{equation}
    PCC(F,R) = \frac{\sum{(F-\bar{F})\cdot (R-\bar{R})}}{\sqrt{\sum{(F-\bar{F})^{2}}\cdot\sum{(R-\bar{R})^{2}}}}
\end{equation}
where $ F$ denotes the fake images generated by PDG and $R$ denotes  the real images in training dataset. $ \bar{F}$  and  $\bar{R}$ represent the mean value of fake and real images, respectively. When the fake images have the same distribution as that of real images, the PCC will become 1, otherwise the PCC will be less than 1. 

\bibliography{scibib.bib}

\bibliographystyle{science}

\section*{Acknowledgments}
\textbf{Funding:} X. Fu acknowledges funding support from Beijing Natural Science Foundation (JQ23021). \textbf{Competing interests:} The authors declare that they have no competing interests. \textbf{Data and materials availability:} All data needed to evaluate the conclusions in the paper are present in the paper and/or the Supplementary Materials.

\section*{Supplementary materials}
Materials and Methods\\
Supplementary Text S1-S5\\
Figs. S1 to S7\\
Tables S1\\
References \textit{(1-4)}

\clearpage

\begin{figure}[!h]
  \centering{
  \includegraphics[width = 1\linewidth]{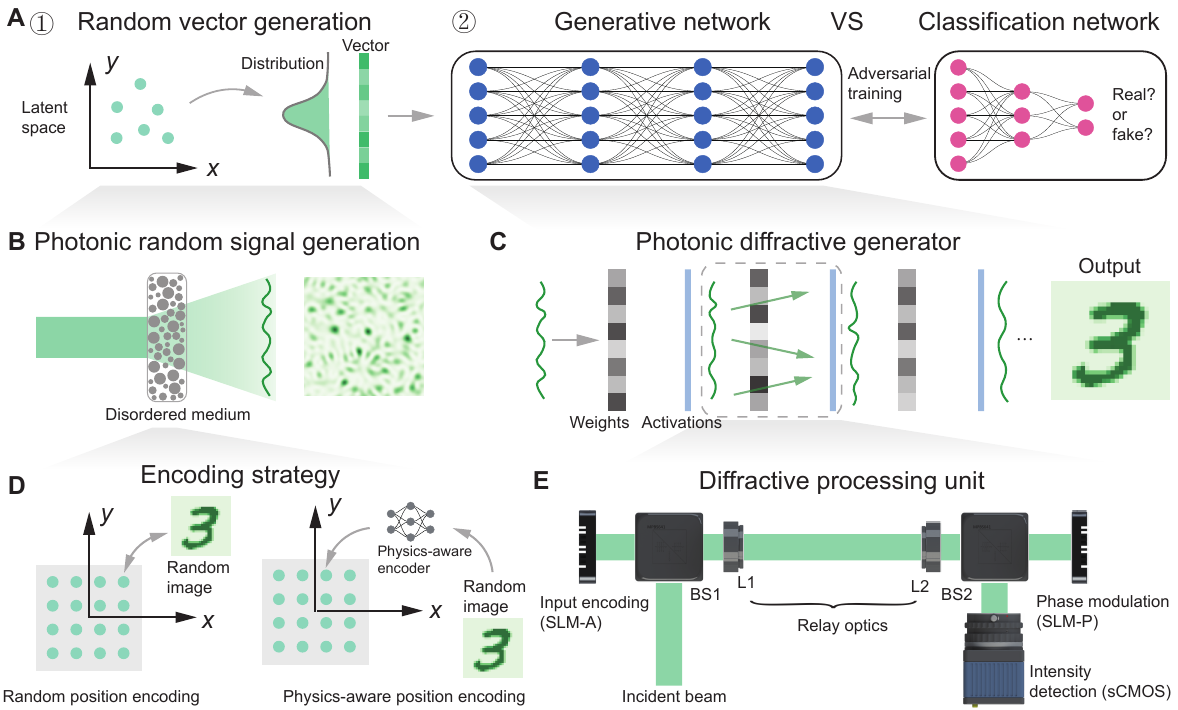}
  } 
    \caption{\noindent\textbf{General workflow of Photonic Diffractive Generator (PDG) } \textbf{(A)} An ordinary GAN architecture is consists of three main components: a random vector sampling source; a generative  network and discriminative network. \textbf{(B)} Photonic random signal source. A CW laser is illuminated at positions of optical disordered media for generating random input signals.  \textbf{(C)} Internal working of the PDG. The input optical signals pass through a well-engineered architecture with unified computing elements called the Diffraction Processing Unit (DPU). \textbf{(D)} Two specific encoding strategies for photonic random signal generation. Left subplot,  images are arbitrarily mapped to different illumination positions without any prior knowledge. Right subplot, random images are preprocessed by a physics-aware encoder, and then mapped with certain illumination positions. \textbf{(E)} Detailed experiment configuration of DPU. L1 and L2, relay lenses; SLM-A amplitude-only spatial light modulator; SLM-P phase-only spatial light modulator. The input laser is first optically encoded by SLM-A, and then undergoes photonic matrix computation, which is generated from phase modulation by SLM-P together with optical free space propagation. The detection process can be then considered as a quadratic non-linear activation function. The entire DPU configuration functions as a single layer perceptron in machine learning.
  }
 \label{Figure1}
\end{figure} 
\begin{figure}
    \centering
    \includegraphics[width = 1\linewidth]{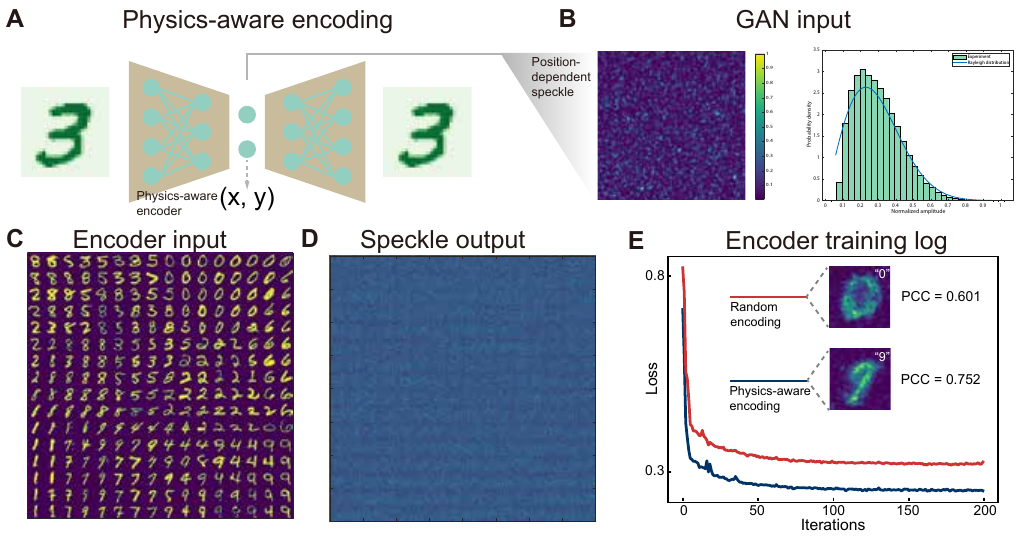}
    \caption{\noindent{\textbf{Physics-aware encoding method.} \textbf{(A)} Schematic graph of  physics-aware encoding.  A digital VAE is tuned in advance upon certain dataset. The encoder of the trained VAE first maps images to two-dimension latent space which corresponds to real-world illumination position ($x,y$). \textbf{(B)} Position-dependent speckles. Left: typical speckle intensity profile. Right: statistical histogram. The intensity of generated speckles follows Rayleigh distribution. \textbf{(C, D)} The encoding results for input images (C) and output speckle (D) are illustrated. \textbf{(E)} Training logs of PDG for two enconding strategies. Both of them are trained under the same environment and hyperparameters. Inset plot: one typical PDG output `0' and `9' for different methods. }}
    \label{fig:enter-label}
\end{figure}
\begin{figure}
    \centering
    \includegraphics[width = 1\linewidth]{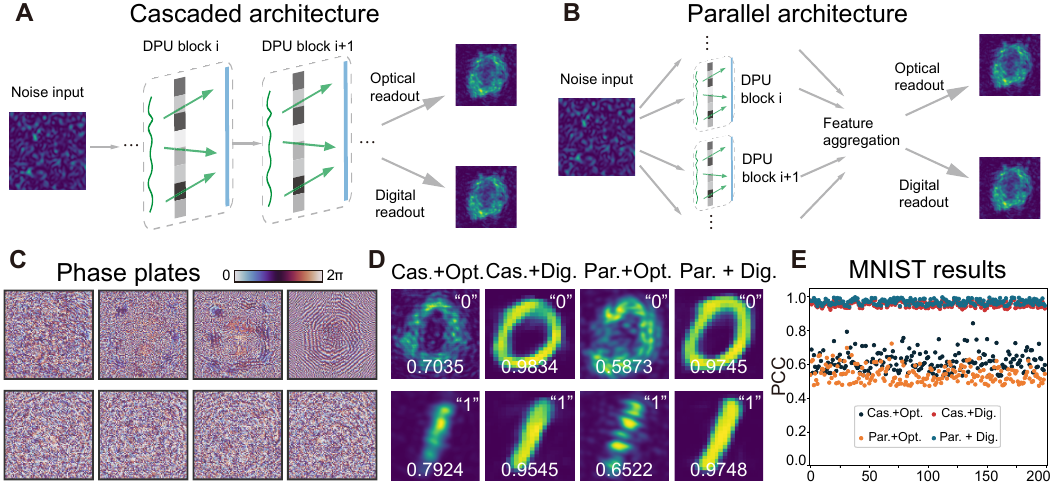} 
    \caption{\noindent{\textbf{PDG architecture and generation of handwritten numbers.} \textbf{(A, B)} Cascaded (A) and parallel (B) architecture of PDG. \textbf{(C)} The designed phase plates for each DPU layer.Upper and lower rows correspond to cascaded and parallel architecture, respectively. \textbf{(D)} Two selected handwritten number (`0' and `1') generation results. Cas.+Opt., cascaded architecture with optical readout; Cas.+Dig., cascaded architecture with digital readout; Par.+Opt., parallel architecture with optical readout; Par.+Dig., parallel architecture with digital readout. \textbf{(E)} Quantitative evaluation of handwritten number generation on 200 chosen experiment results.}}
    \label{fig:enter-label}
\end{figure}
\begin{figure}
    \centering
    \includegraphics[width = 1\linewidth]{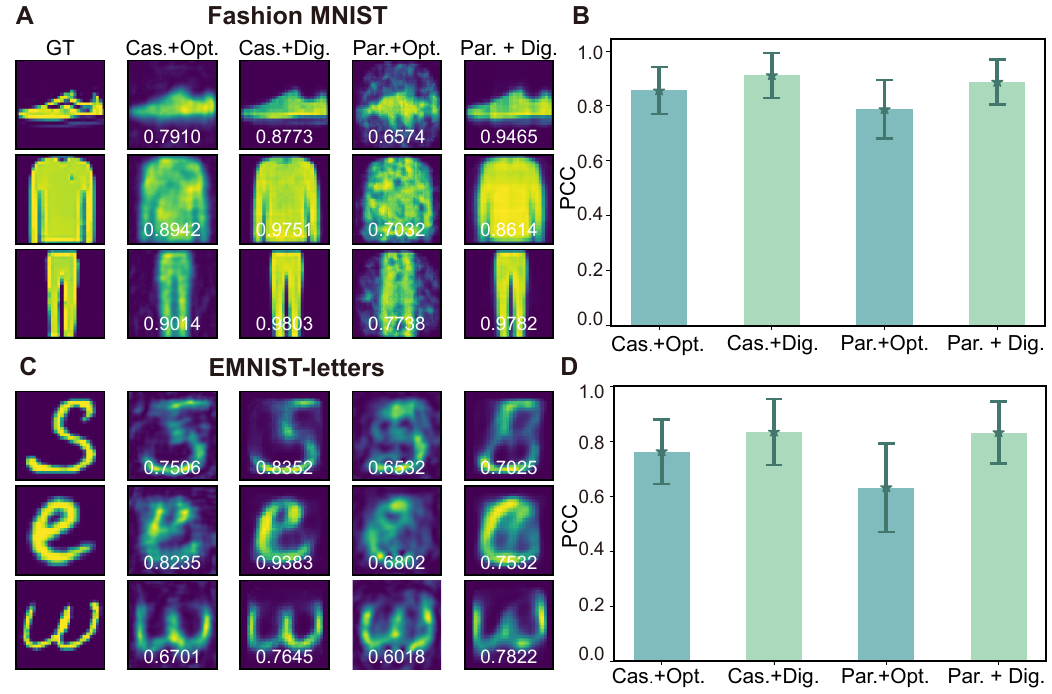} 
    \caption{\noindent\textbf{Simulation results of PDG trained on Fashion MNIST and EMNIST-letter dataset.} \textbf{(A, C)} Three selected simulation results of PDG on Fashion MNIST (A) corresponding to `sneakers', `shirts', `trousers', and on EMNIST (C) corresponding to `S', `e', `w'. \textbf{(B, D)} Statistics analysis of PDG performance over 4096 synthesized images on Fashion MNIST (B) and EMNIST (D). The height of bar represents the mean PCC, while the error bars denotes the standard deviation (STD) of PCC. }
\end{figure}
\begin{figure}
    \centering
    \includegraphics[width = 1\linewidth]{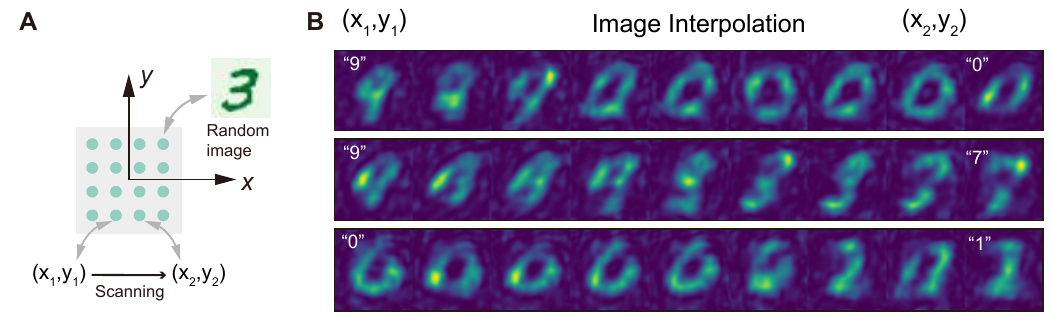} 
\caption{\noindent\textbf{Image interpolation by scanning the illumination position of scattering media} \textbf{(A)} The schematic graph of image interpolation through scanning. \textbf{(B)} Typical results of image interpolation. The scan of illumination position from $(x_1, y_1)$ to $(x_2, y_2)$. The images gradually change from `9' to `0', from `9' to `7' and from `0' to `1'. Except for the first and last images, all intermediate images are new generated ones.}
\end{figure}


\end{document}